# Effect of pressure on superconductivity in doped topological crystalline insulator $Sn_{0.5}In_{0.5}Te$

[1]V. K. Maurya, [2]R. Jha, [1]Shruti, [2]V.P.S. Awana, [1]S. Patnaik[*]

[1]School of Physical Sciences, Jawaharlal Nehru University, New Delhi 110067, India

[2]National Physical Laboratory, New Delhi 110012, India

## Abstract

We report on the impact of hydrostatic pressure on the superconductivity of optimally (Indium) doped SnTe which is established to be derived from a topological crystalline insulating phase. Single crystals of $Sn_{1-x}In_xTe$ were synthesized by a modified Bridgman method that exhibited maximum superconducting $T_c$ of 4.4 K for x= 0.5. Hydrostatic pressure upto 2.5 GPa was applied on the crystals of $Sn_{0.5}In_{0.5}Te$ and electrical resistivity as a function of temperature and pressure was measured. We observed decrease in onset superconducting transition temperature from 4.4 K to 2.8 K on increasing pressure from ambient to 2.5 GPa. The normal state resistivity also decreased abruptly by an order of magnitude at 0.5GPa but for higher pressures, the same decreased marginally. From onset, offset and zero resistivity values, $dT_c/dP$ of ~ -0.6K/GPa was confirmed. The low temperature normal state resistivity followed $T^2$ dependence suggesting Fermi liquid behaviour both for ambient and high pressure data. This increase in metallic characteristics accompanied by normal state Fermi liquid behaviour is in accordance with a "dome structure" for $T_c$ variation with varying carrier concentration.

**Introduction**

Superconductors derived from Topological Insulators (TI) and Topological Crystalline Insulators (TCI) have attracted considerable attention in the recent past [1,2,3]. From theoretical perspective it could bring fruition to the elusive search of Majorana Fermions (MFs) and from the technological point of view it is expected to have significant impact on quantum computation [4,5]. A case in example is indium (In) doped many-valley semiconductor tin-telluride (SnTe) where a maximum superconducting transition temperature of ~4.4 K is reported for x~0.5 in the series $Sn_{1-x}In_xTe$ [6, 3, 7]. Several theoretical studies and detailed angle resolved photo-electron spectroscopic (ARPES) measurements have established SnTe to be a TCI phase due to an underlying mirror symmetry of its crystalline lattice [8,9]. Towards developing an understanding of how these bulk superconductors are different as compared to BCS or cuprate superconductors, in this paper, we report the effect of hydrostatic pressure on the superconducting properties of optimally doped $Sn_{0.5}In_{05}Te$.

The basic premise of measurements under high pressure is that it can effectively tune the electronic and phononic band structure. This has lead to discovery of superconductivity in myriad compounds at high pressure [10-12, 3]. With regard to optimally doped superconductors, such studies could reflect if there is a "dome structure" associated with phase transitions vis á vis carrier concentration that relates to quantum criticality and correlation effects [13]. Further, if pressure could revert to insulating bulk phase, then in principle one can have superconductor - topological insulator interfaces leading to emergence of MFs. Moreover, a dome structure of increasing $T_c$ dependence with varying concentration of Indium upto 50% was indicated earlier in SnTe [6], but the compounds were reportedly multiphase beyond x = 0.5 and therefore high pressure studies are essential to elucidate full phase diagram of $Sn_{1-x}In_xTe$.

In particular, topological surface states have been prominently investigated in $Bi_2Se_3$ and $Bi_2Te_3$ which have been driven to superconducting state with application of external pressure [10,11, 12]. In undoped topological insulators, superconductivity is achieved at relatively high pressure. For example, $Bi_2Se_3$ shows a turnover from semiconducting to metallic behaviour at ~ 8 GPa accompanied by a structural phase transition [14]. Superconductivity appears in $Bi_2Se_3$ at ~13.5 GPa at a transition temperature of 0.5K, which gradually increases to a maximum of 7 K on increasing pressure upto 30GPa. At higher pressures, the $T_c$ remains almost constant upto 50 Gpa [10]. $Bi_2Te_3$ is another topological insulator, for which superconductivity is reported around 3 K under 3 GPa pressure that increased to 8 K at 15 GPa and for further higher pressures the $T_c$ exhibited a decreasing trend [12]. Like $Bi_2Se_3$, $Bi_2Te_3$ also undergoes several structural transformations from rhombohedral (R-3m) phase to monoclinic (C2/m) at 3GPa, to monoclinic (C2/c) at 8 GPa and finally the BCC Im-3m structure at above 16 GPa [12].

Superconductors derived from topological insulators through intercalation have also been subjected to various pressure studies [15]. $Cu_xBi_2Se_3$ is a well-known topological insulator based superconductor with a maximum $T_c$ of around 3.8 K [15]. Point contact spectra on the surface of Cu intercalated $Bi_2Se_3$ exhibited signs of unconventional superconductivity [16]. On applying pressure on $Cu_xBi_2Se_3$ a gradual decrease in the superconducting transition temperature is reported and as pressure is increased further superconductivity disappeared at ~6.3 GPa [17]. With regard to TCI systems, a first principle study on SnTe predicts a maximum superconducting $T_c$ = 7.16 K by pressure tuning the electron-phonon coupling parameters in the body center cubic phase (Pm-3m) [18]. In this paper, we focus on impact of pressure on superconducting and normal state conduction of recently discovered optimally doped TCI superconductor $Sn_{0.5}In_{0.5}Te$. We found that superconducting transition temperature ($T_c$) decreased monotonically with pressure (~ 0.6K/GPa) and the normal state resistivity also decreased by an order of magnitude at 0.5GPa. Such behaviour finds surprising resemblance with the curious case of over-doped cuprates.

**Experimental methods**

Single crystals of $Sn_{0.5}In_{0.5}Te$ were prepared by a modified Bridgman method. A series of compound with varying indium concentration were prepared and optimum superconducting $T_c$ was achieved for $Sn_{0.5}In_{0.5}Te$ [7]. We studied electrical resistivity at high pressure on this composition. Single crystals were obtained by melting stoichiometric amounts of high purity elemental powder of Sn (99.99%), Te (99.999%) and shots of In(99.99%) at 900$^o$C for 5 days in sealed evacuated quartz tubes. Intermittent shaking was performed for the homogeneity of melt sample. Sample was cooled to 770$^o$C over a period of 72 h followed by annealing at 770$^o$C for 48 hours. Silvery - shiny single crystals were cleaved along z axis. X-ray Diffraction was carried out on the powdered samples by *RIGAKU* powder X- ray Diffractometer (Miniflex 600) [7].

The pressure dependent resistivity measurements were performed in Physical Property Measurements System (*PPMS*-14T, *Quantum Design*) using HPC-33 Piston type pressure cell with Quantum design DC resistivity Option. Hydrostatic pressures were generated by a BeCu/NiCrAl clamped piston-cylinder cell. The sample was immersed in a fluid (Daphne Oil) with pressure transmitting medium of Fluorinert in a Teflon cell. Annealed Pt wires were affixed to gold-sputtered contact surfaces on each sample with silver epoxy in four-probe configuration.

**Results and discussion:**

The powder XRD pattern of $Sn_{0.5}In_{0.5}Te$ is shown in Figure 1. It confirms pure phase synthesis in agreement with reference data from JCPDF (No. 089-3974). The specimen crystallizes in rock-salt structure with space group Fm-3m. The calculated lattice parameter is **a** = 6.265 Å and the cell volume is 245.65 Å$^3$. In the inset as grown crystal flakes are shown. The electrical resistivity as a function of temperature ($\rho$-T) for $Sn_{0.5}In_{0.5}Te$ at ambient pressure is shown in Figure 2. The inset shows resistivity upto room temperature. We mark $T_c^{onset}$ by the intersection of the two extrapolated lines, one corresponding to superconducting transition line and the other is an extended normal state resistivity line. Similarly $T_c^{offset}$ is indicated by intersection of transition line and a zero resistivity line. We define $T_c^{zero}$ as temperature where zero resistivity state was achieved. This is schematically shown in Figure 2. From Figure 2, The values of $T_c^{onset}$, $T_c^{offset}$ and $T_c^{zero}$ for single crystal $Sn_{0.5}In_{0.5}Te$ are found to be 4.4 K, 4.1 K and 3.6 K respectively. The superconducting transition is sharp with a transition width of ~0.3 K.

The resistivity Vs. Temperature behaviour near superconducting transition ($T_c$) for varying pressure is shown in Figure 3(a). For clarity, the data for ambient pressure are not included as it is an order of magnitude higher. It is seen that both superconducting transition temperature and normal state resistivity decrease with increasing pressure. It can be seen that at maximum pressure of 2.5 GPa, the $T_c^{onset}$ is decreased to 2.8 K from 4.4 K (ambient pressure), while the $T_c^{offset}$ and $T_c^{zero}$ are decreased to 2.6 K and 2.3 K from 4.1 K and 3.6 K respectively. The $T_c^{onset}$, $T_c^{offset}$ and $T_c^{zero}$ for $Sn_{0.5}In_{0.5}Te$ superconductor at intermediate pressure are summarized in Figure 3(b). It seems that decrease in the $T_c$ is about linear for two markers (onset, offset) of superconducting transition. The negative coefficients of $T_c$ suppression with pressure for three markers ($T_c^{onset}$, $T_c^{offset}$ $T_c^{zero}$) are estimated to be -0.66 K/GPa, -0.61 K/GPa and -0.57 K/GPa respectively that yield average $dT_c/dP$ of -0.6 K/GPa for $Sn_{0.5}In_{0.5}Te$ superconductor.

In Figure 4(a) we compare resistivity versus temperature (ρ-T) for $Sn_{0.5}In_{0.5}Te$ taken at ambient pressure and applied pressures in extended temperature range of up to 250 K. This is done to visualise the impact of hydrostatic pressure on the normal state conduction of $Sn_{0.5}In_{0.5}Te$. We see that at applied pressure 0.5 GPa the normal resistivity (resistivity just above the transition) decreases abruptly by nearly an order of magnitude. Quantitatively, a decrease of about 7.8 times (659.76 μΩ-cm to 84.09 μΩ-cm) is observed. With further increase in pressure, while the normal state resistivity continues to decrease, the rate of change with pressure decreases substantially. Over all the change in resistivity with temperature shows metallic behaviour at both ambient and applied pressures of up to 2.5 GPa and clearly the metallic behaviour increases with higher pressure. This is in contrast to $Cu_xBi_2Se_3$ where ρ ($T_c$) increased by 7 times at 2.31 GPa. For low carrier density superconductors the BCS equation relates superconducting transition temperature with carrier concentration; $T_c \sim \theta_D \exp(-1/N(E_F) V_0)$ where $\theta_D$ is the Debye temperature, $V_0$ is the electron-phonon coupling calliper and density of state $N(E_F) \sim m^* n^{1/3}$ which is a product of effective mass $m^*$ and carrier concentration n. A decreasing carrier concentration (increasing normal state resistivity) with pressure can explain decreasing superconducting $T_c$. But we find that normal state resistivity for $Sn_{0.5}In_{0.5}Te$ decreased many-fold with pressure. Thus the controlling parameter for $T_c$ suppression mechanism seems to be decreasing effective mass $m^*$ in $Sn_{0.5}In_{0.5}Te$ under a simplistic S wave correlation [7]. In Figure 4(b) we show the change in the resistivity just above the transition temperature and at 250 K for ambient and various applied pressures. We can see that both decrease very fast on application of 0.5 GPa but for higher pressure, the change is relatively much less. The RRR (residual resistivity ratio between resistivity at 250 K temperature and temperature just above the transition) remains ~1.25 from ambient pressure to 2.5 GPa. This indicates that within experimental error, the impurity band contribution to the normal state conduction mechanism in $Sn_{0.5}In_{0.5}Te$ remains majorly unaffected by applied pressure.

We note that both in high $T_c$ cuprates and in pnictide superconductors, the transition temperature varies in a so called dome structure as a function of doping concentration. In the overdoped region, the normal state resistivity is well characterized by Fermi liquid behaviour. Towards studying appropriateness of Fermi liquid theory (negligible electronic correlations), in Figure 5 we plot the resistivity versus $T^2$ curves in the temperature range 11 K to 30 K for $Sn_{0.5}In_{0.5}Te$. In this theory, particularly with regard to heavy fermion systems, the strong interaction between charge carriers is replaced by weakly correlated quasi-particles with high effective mass. In Figure 5, the resistivity data taken at various pressure are fitted to the equation $\rho = \rho_0 + AT^2$ where $\rho_0$ relates to impurity scattering and the coefficient A relates to square of the effective mass of quasi-particles. The calculated values for $\rho_0$ and A are tabulated in Table 1. We find that for the pressure 0, 0.5, 1, 1.5, 2, 2.5 GPa the data follow $T^2$ behaviour in the temperature range 11 K to 30 K and the curves deviate from linearity above 30 K. In the inset of Fig. 5 we plot pressure dependence of coefficient A which indicates a decreasing trend with increasing pressure. This is suggestive of weakening correlation between charge carriers in the over-doped region. But the origin of this correlation phenomenon in the semiconducting parent SnTe needs to be ascertained.

**Conclusion:**

In summary, we have prepared single crystals of $Sn_{0.5}In_{0.5}Te$ and applied pressure up to 2.5 GPa to check its superconducting properties under pressure. This is an optimally doped specimen derived from a topological crystalline insulating phase. We found that superconducting $T_c$ (onset) decreased with pressure from 4.4 K (ambient) to 3.8 K (2.5GPa). This suppression of superconducting transition temperatures was found to be almost monotonic with pressure and the overall $dT_c/dP$ is estimated to be -0.6K/GPa. Fermi liquid behaviour was indicated in the temperature range 11 K to 30 K and we found that normal state resistivity of sample varies as a function of $T^2$ with increasing pressure. Such systematic decrease with $T_c$ with increasing metallicity and normal state $T^2$ behaviour is reminiscent of over-doped high $T_c$ cuprates.

**Acknowledgement:**

SP thanks DST- FIST and VPSA thanks DAE-SRC for supporting research infrastructure at JNU and NPL respectively. VM, Shruti and RJ acknowledge financial support through senior research fellowships from UGC-BSR, UGC, and CSIR respectively. SP thanks G Baskaran and Brijesh Kumar for discussions on the experimental data.

**Table Caption**

The values for $\rho_0$ and A at various pressure for the equation $\rho = \rho_0 + AT^2$

**Figure captions**

Fig.1 Powder XRD pattern of crystals of $Sn_{0.5}In_{0.5}Te$. In the inset picture of crystals is shown.

Fig.2: Resistive superconducting transition of $Sn_{0.5}In_{0.5}Te$ at ambient pressure. $T_c^{onset}$, $T_c^{offset}$ $T_c^{zero}$ are indicated by arrows.

Fig.3(a) : Resistive superconducting transition of $Sn_{0.5}In_{0.5}Te$ at different pressure from 0.5 GPa to 2.5 GPa. An unambiguous decrease in superconducting transition temperature is seen.

Fig.3(b) : Variation in superconducting $T_c$ at various pressure for $Sn_{0.5}In_{0.5}Te$. We can see a negative pressure coefficient ($dT_c/dp$) for all transitions $T_c^{onset}$, $T_c^{offset}$ and $T_c^{zero}$.

Fig.4(a) : Resistivity upto 250 K for $Sn_{0.5}In_{0.5}Te$. A large decrease in resistivity is seen on application of 0.5 GPa. For higher pressure, resistivity decreases at a smaller rate. No significant variation in Residual Resistivity Ratio (RRR) is observed.

Fig.4(b) : Magnitude of electrical resistivity at 250 K and at temperature close to $T_c$ is plotted as a function of pressure. Sharp decrease in magnitude is observed at 0.5GPa followed by almost linear decrease in the resistivity.

Fig.5 : Resistivity Vs. $T^2$ plot for pressure 0, 0.5, 1, 1.5, 2, 2.5 GPa. The solid lines are linear fit to the equation $\rho = \rho_0 + AT^2$.

**Table 1**

| Pressure (GPa) | 0 | 0.5 | 1 | 1.5 | 2 | 2.5 |
|---|---|---|---|---|---|---|
| $\rho_0(\mu\Omega\text{-cm})$ | 669.96 | 82.5 | 68.07 | 69.1 | 67.2 | 60.8 |
| $A(\mu\Omega\text{-cm/T}^2)$ | $8.49\times10^{-3}$ | $3.71\times10^{-3}$ | $2.75\times10^{-3}$ | $2.49\times10^{-3}$ | $5.25\times10^{-4}$ | $1.58\times10^{-4}$ |

**Figure 1**

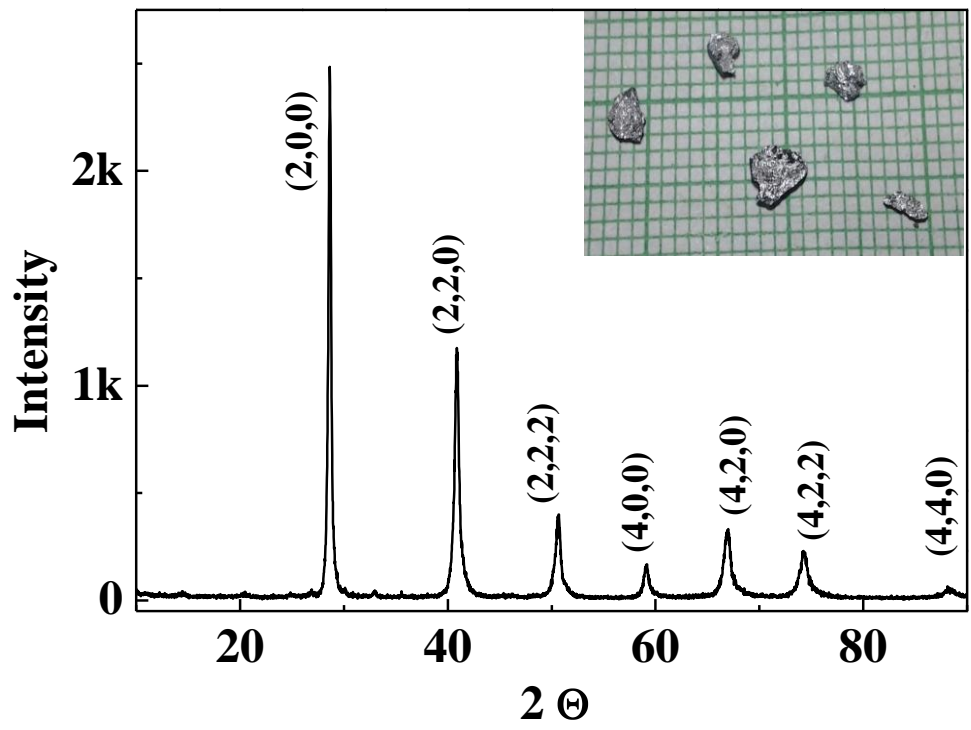

**Figure 2**

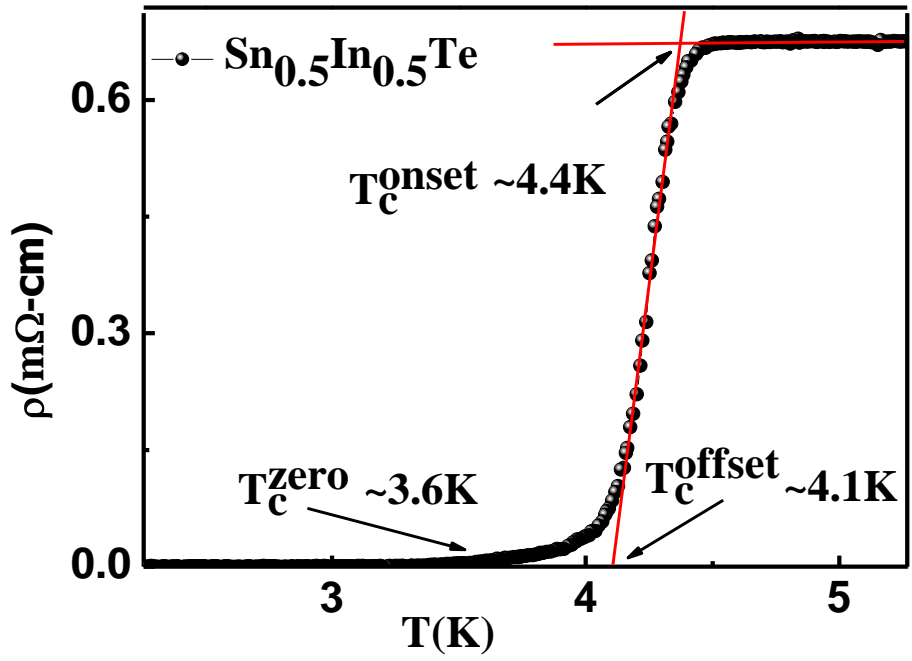

**Figure 3(a)**

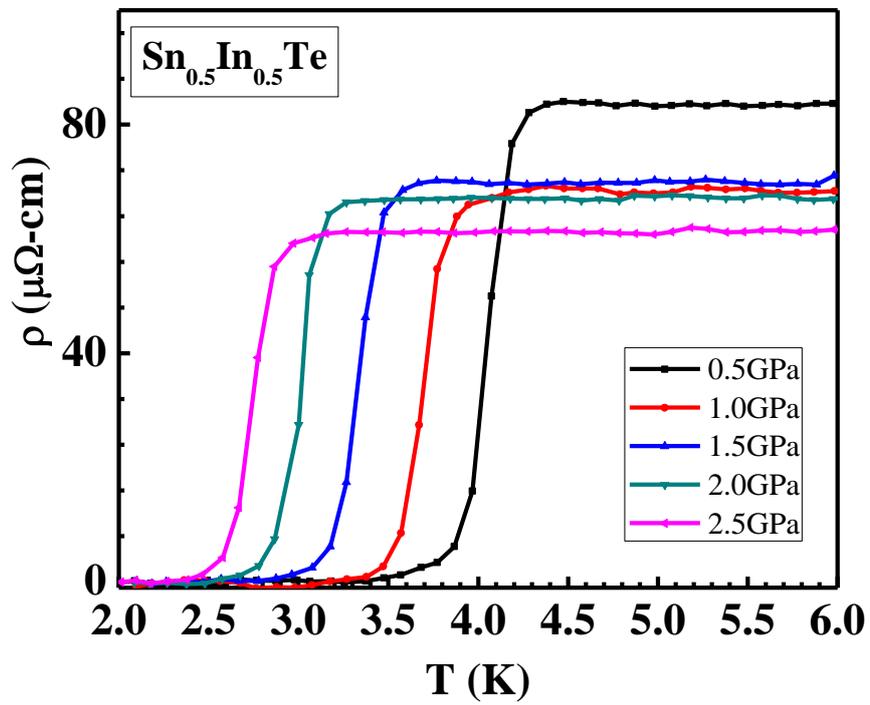

**Figure 3(b)**

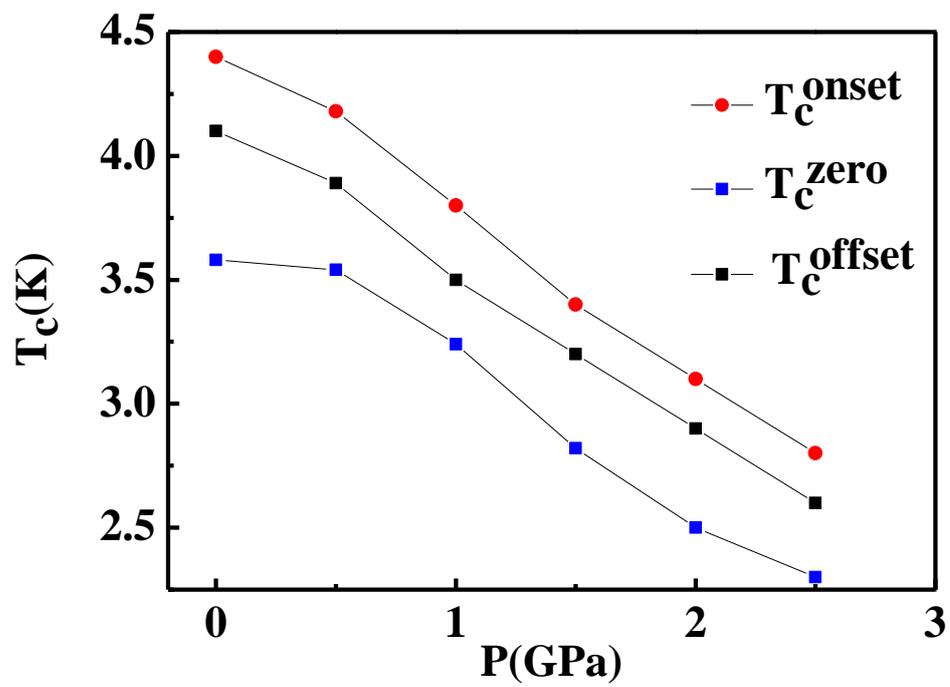

**Figure 4(a)**

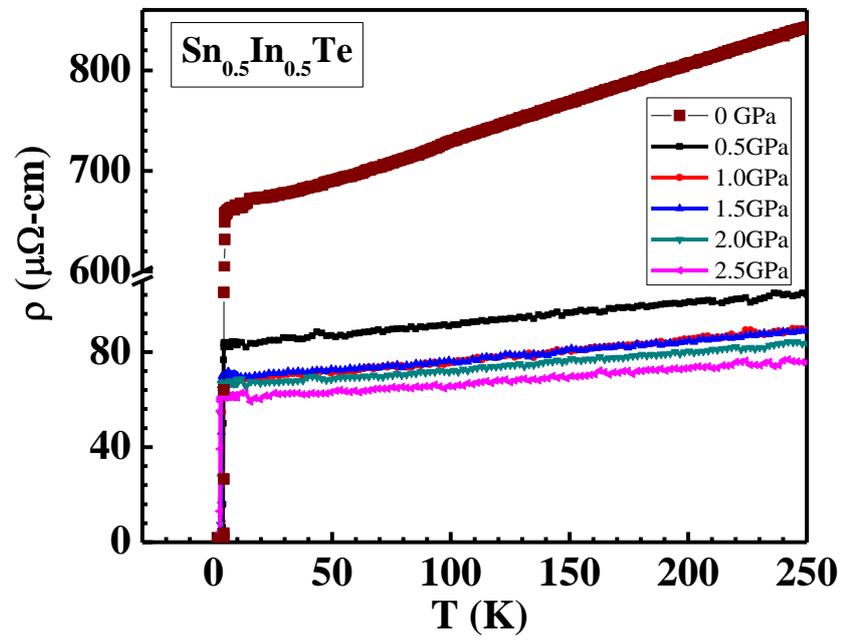

**Figure 4(b)**

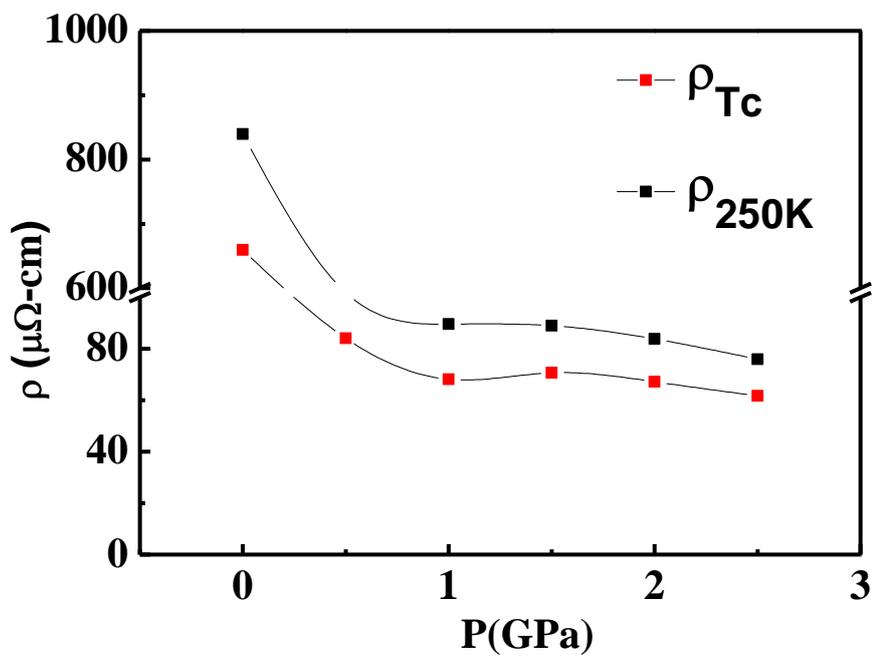

**Figure 5**

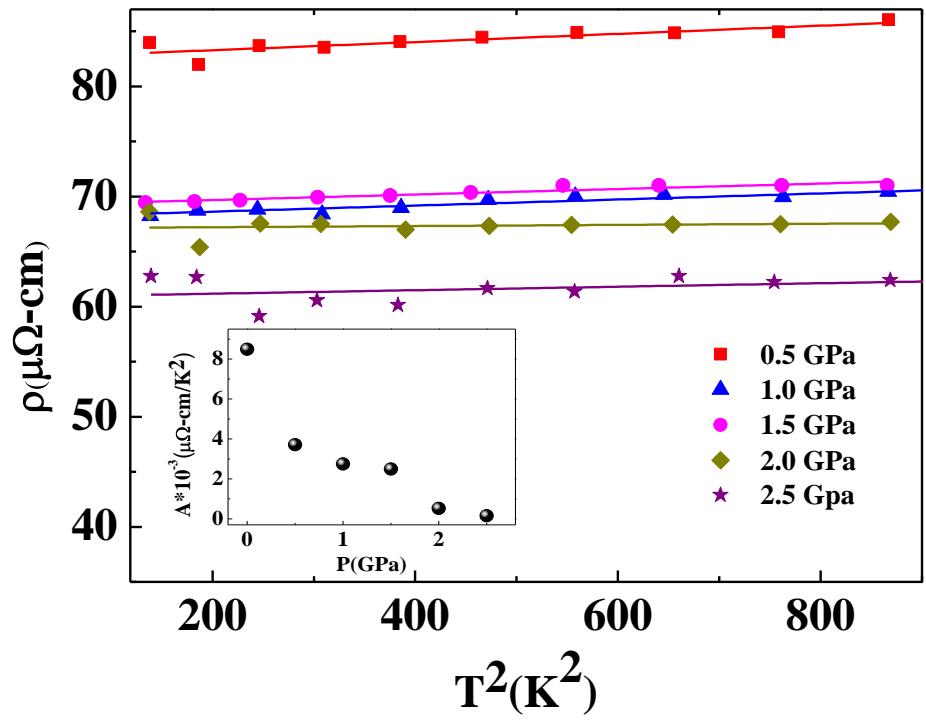